\documentclass[prl,aps,twocolumn,showpacs]{revtex4}
\usepackage{graphicx,epsfig,amsmath,amsfonts,bm}

\begin{document}

\title{Exact results for anomalous transport in one-dimensional Hamiltonian systems}

\author{Henk van Beijeren}
\email{H.vanBeijeren@uu.nl}
\affiliation{Institute for Theoretical Physics, Utrecht University, Leuvenlaan 4, 3584 CE, Utrecht, The Netherlands}
\pacs{05.20.Jj  05.60.Cd  05.70.Ln  47.10.-g  62.65.+k  63.22.-m  66.25+g}

\begin{abstract}
\noindent
Anomalous transport in one-dimensional translation invariant Hamiltonian systems with short range interactions, is shown to belong in general to the KPZ universality class. Exact asymptotic forms for density-density and current-current time correlation functions and their Fourier transforms are given in terms of
the Pr\"ahofer-Spohn scaling functions, obtained from their exact solution for the polynuclear growth model. The exponents of corrections to scaling are found as well, but not so the coefficients. Mode coupling theories developed previously are found to be adequate for weakly nonlinear chains, but in need of corrections for strongly anharmonic interparticle potentials. A simple condition is given under which KPZ behavior does not apply, sound attenuation is only logarithmically superdiffusive and heat conduction is more strongly superdiffusive than under KPZ behavior.
\end{abstract}

\date{\today}

\maketitle

Since the discovery by Alder and Wainwright\cite{alderw} of long-time tails in the Green-Kubo current-current time correlations, such as the velocity autocorrelation function it has been clear that transport in one and two dimensional Hamiltonian systems must be anomalous in most cases. One-dimensional systems have been 
studied extensively in the past decades, both by mode coupling techniques\cite{delfini,delfini2,lietal,leedadswell} and dynamical scaling\cite{rama}, and also by computer simulations\cite{delfini,delfini2,lietal,leedadswell,grassberger}.
Most studied are the exponents $\alpha$ describing the divergence of 
the coefficients of heat conduction and sound damping with system size $L$ as $L^{\alpha}$, and $\delta$ describing the power law $t^{-(1-\delta)}$ by which the corresponding current-current time correlation functions decay. For both exponents various values have been proposed, with $\delta
=1/3$ for both heat conduction and sound attenuation and $\alpha=1/3$ for heat conduction but 1/2 for sound attenuation being the most common ones in recent publications.

Here I will argue that for generic Hamiltonian systems the long time behavior of the dynamics can be obtained {\em exactly} in terms of the scaling functions obtained by Pr\"ahofer and Spohn\cite{PS} for the polynuclear growth model, which is in the KPZ universality class. The values of  $\delta$ and $\alpha$ mentioned above are confirmed. But also the {\em coefficients} of size dependent transport coefficients and long-time current-current correlation functions are obtained exactly, as well as the scaling functions describing, among other things the asymptotic behaviors of the various density-density time correlation functions and their Fourier transforms. These results hold in all generality, for generic short ranged 1d Hamiltonians, from weakly anharmonic chains up to mixtures of hard points. They establish a rare example of exact results that may be obtained for non-integrable Hamiltonian systems out of equilibrium. In addition the special conditions under which such systems do not belong to the KPZ universality class will be formulated simply and sharply, together with the consequences for long time and short wave length behavior. 
 
More specifically, I will discuss classical one-dimensional N-particle systems described by a translation invariant Hamiltonian with short range interactions and periodic boundary conditions. 
Following one of the ground-laying papers by Ernst, Hauge and Van Leeuwen\cite{ehvl} I will assume
that all slow variables of relevance for the long time behavior of hydrodynamics and related time correlation functions are the long-wave length Fourier components of the densities of conserved quantities, i.e.\  particle number, momentum and energy, plus products of these. This is a crucial assumption. It is not satisfied for most exactly solvable models, which have additional slow modes, such as solitons\cite{toda}. For one- and two-dimensional systems the method of EHvL has to be generalized somewhat: instead of assuming that the time correlation functions of hydrodynamic modes decay exponentially with time, one has to write down the mode coupling equations as a set of coupled nonlinear equations for these correlation functions that must be solved self-consistently\cite{delfini,delfini2,lietal}.

EHvL define the hydrodynamic modes, to leading order in the wave number $k$ as linear combinations of the Fourier transforms of the microscopic densities of particles, momentum and energy\footnote{The sign convention used here for the spatial Fourier transforms is opposite to that used by Pr\"ahofer and Spohn, but the same as that used by  Ernst, Hauge and Van Leeuwen\cite{ehvl}. Since the scaling functions obtained by Pr\"ahofer and Spohn are even in $k$ this is of no severe consequence.}, $\rho^{\mu}(k,t)=\sum_{j=1}^N M^{\mu}_j \exp(-ikx_j)-\delta_{k0}\langle \hat{M}(k=0)\rangle$, with $M^{\mu}_j=1,p_j,e_j$ for the particle density $n(k,t)$, the momentum density $g(k,t)$ and the energy density $e(k,t)$ respectively 
\footnote{Note that the energies and momenta of the particles are localized at the actual positions of the particles. For chains on which particles cannot pass each other, a natural alternative is localizing these densities at the average  positions of the particles\cite{delfini,delfini2,lietal}.
However, one can show that both approaches are equivalent, at least so in their predictions of the long time dynamics\cite{tbp}}. 
The hydrodynamic modes are two sound modes\footnote{I use $\sigma=\pm1$  for right respectively left moving sound modes, rather than positive respectively negative frequency, as is conventional.} $a_1(k,t)$ and $a_{-1}(k,t)$ and a heat mode $a_H(k,t)$, given respectively, to leading order in $k$ by
\begin{align}
&a_{\sigma}(k,t)=\left(\frac{\beta}{2\rho}\right)^{1/2}\left(c_0^{-1}p(k,t)+\sigma g(k,t)\right),\label{asigma}\\
&a_H(k,t)=\left(\frac{\beta}{nTC_p}\right)^{1/2}(e(k,t)-hn(k,t)).\label{aH}
\end{align}
Here, $\sigma$=$\pm1$, $T$ is the equilibrium temperature, $n$ the equilibrium number density and $\rho$=$nm$;
$C_p$=$T(\partial s/\partial T)_p$ is the specific heat per particle at constant pressure $p$, with $s$ the equilibrium entropy per particle; $c_0$=$(\partial p/\partial \rho)_s^{1/2}$ is the adiabatic sound velocity in the limit of zero wave number and $h$ is the equilibrium enthalpy per particle.
Furthermore,
\begin{align}
p(k,t)&=(\partial p/\partial e)_n e(k,t) +(\partial p/\partial n)_e n(k,t),\label{asound}\\
&=\frac{\gamma-1}{\alpha T}e(k,t) +(\partial p/\partial n)_e n(k,t),\nonumber
\end{align} 
where $\gamma=C_p/C_v$ is the specific heat ratio and $\alpha=-n^{-1}\left(\partial n/\partial T\right)_p$ the thermal expansion coefficient.
The allowed values of $k$ are of the form
$k=\frac{2\pi n} L$. To leading order in $k$ the hydrodynamic modes are normalized under the inner product $(f,g)=\frac 1 L \langle f^*g\rangle,$ with $\langle \rangle$ a grand canonical equilibrium average.

The time correlation functions of the hydrodynamic modes satisfy linear equations involving memory kernels,
viz.\
\begin{align}
&\frac{\partial \hat{S}_{\sigma}(k,t)}{\partial\ t}=-i\sigma c_0 k\hat{S}_{\sigma}(k,t)-k^2 \int_0^t d\tau \hat{M}_{\sigma}(k,\tau)\hat{S}_{\sigma}(k,t-\tau),\label{soundmem}\\
&\frac{\partial \hat{S}_{H}(k,t)}{\partial\ t}=-k^2 \int_0^t d\tau \hat{M}_{H}(k,\tau)\hat{S}_{H(k,t-\tau)}.\label{heatmem}
\end{align}
Here $\hat{S}_{\sigma}(k,t)=(a_{\sigma}(k,0),a_{\sigma}(k,t))$ etc. 
The memory kernels may be expressed through a diagrammatic mode coupling expansion as a sum of irreducible skeleton diagrams\cite{skeleton}. These consist of propagators, representing stationary density correlation functions $\hat{S}_{\zeta}(\ell,t_{\alpha})$, and vertices representing the coupling of one propagator $\hat{S}(\ell,t_{\alpha})$ to two propagators $\hat{S}_{\mu}(q,t_{\alpha'})$ and $\hat{S}_{\nu}(\ell-q,t_{\alpha''})$, with coupling strength $\ell W_{\zeta}^{\mu\nu}$. For the long time dynamics only a few of these 27 couplings are important; only couplings to two sound modes of the same sign 
or to two heat modes may give rise to long-lived perturbations, all other combinations of pairs of modes rapidly die out through oscillations. From EHvL\cite{ehvl} the relevant non-vanishing coupling strengths to leading order in $k$ can be obtained as
\begin{align}
W_{\sigma}^{\sigma\sigma}&=\frac{\sigma}{(2\rho\beta)^{1/2}c_0}\left(\frac{\partial c_0n}{\partial n}\right)_s\label{wsss}\\
W_{\sigma}^{-\sigma-\sigma}&=\frac{\sigma}{(2\rho\beta)^{1/2}}\left[\frac 1{c_0}\left(\frac{\partial c_0n}{\partial n}\right)_s-2\frac{\gamma-1}{\alpha T}\right]\label{wsss'}\\
W_{\sigma}^{HH}&= 
\frac{-\sigma(\gamma -1)
}{(2\rho\beta )^{1/2}nC_p}\left(\frac{\partial\left[\frac{n  C_p}{\alpha}\right]}{\partial T}\right)_p\label{wshh}\\
W_{H}^{\sigma\sigma}&=\frac{\sigma k_B^{1/2}c_0}{(n C_p)^{1/2}}.\label{whss}
\end{align}

Now a central observation is the following: due to the first term on the right-hand side of Eq.\ (\ref{soundmem}) the sound-sound correlation functions will have their weights centered around the positions $x(t)=x(0)\pm c_0t$, in other words, these functions will assume the forms $\hat{S}_{\sigma}(k,t)=\exp(-i\sigma c_0kt)\hat{\Sigma}_{\sigma}(k,t)$, with $\hat{\Sigma}_{\sigma}(k,t)$ to a first approximation real non-oscillating functions.
 As a consequence the mode coupling contributions to $\hat{M}^{\sigma}$ are dominated by those diagrams in which all vertices are of the type $V_{\sigma}^{\sigma\sigma}$ (but, only in the limit $k=0$ diagrams in which the first and last vertex are of type $V_{\sigma}^{-\sigma-\sigma}$ and all the other ones of type $V_{-\sigma}^{-\sigma-\sigma}$ also contribute to the leading order). All other contributions for at least some time will oscillate out of phase with the angular frequency $\sigma c_0k$ of the sound mode under consideration. The remaining contributions, especially so if described in a coordinate frame comoving at the speed of sound, can be identified with the terms in a similar mode coupling expansion for the fluctuating Burgers equation\cite{mcburgers},
 \begin{align}
 &\frac{\partial \rho(x,t)}{\partial t}=\frac{\kappa} 2\frac{\partial \rho^2}{\partial x} +\frac D{A}\frac{\partial^2 \rho}{\partial x^2} + \frac{\partial \eta}{\partial x},
 \label{FB} 
\end{align}
with $A$=$\hat{S}_B(0,0)$, with the density-density time correlation function $\hat{S}_B(k,t)$ defined, in the limit $L\to\infty$ as
\begin{align} 
\hat{S}_B(k,t)=\int_{-\infty}^{\infty} dx e^{-ikx} S_B(x,t)\equiv \int_{-\infty}^{\infty} dx e^{-ikx}\langle\rho(0,0)\rho(x,t)\rangle\nonumber
\end{align}
and $\eta(x,t)$ representing gaussian white noise with $\langle\eta(x,t)\eta(x',t')\rangle$=$2D\delta(x-x')\delta(t-t')$. The brackets denote an average over the stationary distribution of the density field.
This is similar to the hydrodynamic equations, but simpler because there is only one conservation law. The function $\hat{S}_B(k,t)$ satisfies an equation similar to Eqs.\ (\ref{soundmem},\ref{heatmem}), of the form
\begin{align}
 &\frac{\partial\hat{S}_B(k,t)}{\partial t}=-k^2\int_0^t d\tau \hat{M}_B(k,\tau)\hat{S}_B(k,t-\tau).
\label{burgmem}
\end{align}
The mode coupling expansion for this memory kernel has exactly the same structure as the set of dominant terms for the sound mode memory kernel; all propagators correspond to the same type of correlation function and all vertices carry the same weight factor $W$, in the case of the Burgers equation given by $W_B=\kappa\sqrt{A} $.

From their exact solution of the polynuclear growth model\cite{PS} Pr\"ahofer and Spohn obtained exact expressions for the long time, respectively small frequency behavior of the function $\hat{S}_B(k,t)$ and its temporal Fourier transform $\tilde{S}_B(k,\omega)$. In the infinite system limit, $L\to\infty$ these are of the form\cite{sasa}
\begin{align}
&\hat{S}_B(k,t)=A\hat{f}_{PS}\left((2A\kappa^2t^2)^{1/3}k\right)\label{hatfPS}\\
&\tilde{S}_B(k,\omega)=\sqrt\frac{A}{2\kappa^2|k|^3}\mathring{f}_{PS}\left(\frac{\omega}{(2A\kappa^2)^{1/2}|k|^{3/2}}\right)\label{mathringfPS},
\end{align}
with the functions $\hat{f}_{PS}$ and 
$\mathring{f}_{PS}$  defined in Eqs.\ ((5.3) and (5.7) of Ref.\cite{PS}. From Eq.\ \ref{burgmem} one may obtain expressions for the memory kernel in terms of these scaling functions. For the full Fourier transform one obtains
\begin{align}
\tilde{M}_B(k,\omega)&=\sqrt{ 2A\kappa^2}M_{PS}\left(k,\frac{\omega}{\sqrt{2A\kappa^2}}\right),\label{Mkomega}
\end{align}
with
\begin{align}
\tilde{M}_{PS}(k,\omega)&=\frac {i\omega}{k^2} + \left(\sqrt{k}\mathring{f}^+_{PS}\left(\frac{\omega}{|k|^{3/2}}\right)\right)^{-1},
\label{mem}
\end{align}
where $\mathring{f}_{PS}^+(w)\equiv\int_0^{\infty} d\tau\exp(iw \tau)\hat{f}_{PS}(\tau^{2/3})$.

The corresponding expressions for the long time behavior of the sound modes at non-vanishing $k$ are
\begin{align}
&\hat{S}_{\sigma}(k,t)=\exp(-i\sigma c_0kt)\hat{f}_{PS}\left((\sqrt{2}V_st)^{2/3}k\right),
\label{Ssigma}\\
&\hat{M}_{\sigma}(k,t)=2V_s^2\exp(-i\sigma c_0kt)\hat{M}_{PS}(k,\sqrt{2}V_st),
\label{Msigma}
\end{align}
with $V_s=W_{\sigma}^{\sigma\sigma}$.

Next, I consider the wave number dependent sound damping constant $\Gamma(k)\equiv2\tilde{M}_{\sigma}(k,0)$ and define the sound currents as 
\[\hat{J}_{\sigma}(k,t)=\left(\frac{\beta}{2\rho}\right)^{1/2}\left[\sigma\hat{J}_l(k,t)+\frac {\gamma -1} {\alpha T c_0}\hat{J}_H(k,t) \right],\]
where $\hat{J}_l(k,t)$ and $\hat{J}_H(k,t)$ are the longitudinal current and the heat current\cite{ehvl}, 
denoted by EHvL as $J_l$ and $J_{\lambda}$ respectively. Eq.\ (5.11) of Ref.\cite{PS} can now be used to obtain the leading small-$k$ behavior of $\Gamma(k)$ and long time behavior of $\langle\hat{J}_{\sigma}(0,0)\hat{J}_{\sigma}(0,t)\rangle$ as
\begin{align}
&\Gamma(k)=\frac {8} {19.444} \sqrt\frac{2V_s^2}{|k|}\label{gammak}\\
&\frac 1 L \langle \hat{J}_{\sigma}(0,t)\hat{J}_{\sigma}(0,0))\rangle= \frac{2.1056[V_s^2 +V_{s'}^2]}{2\sqrt{3}\Gamma_E(1/3)}\left(\frac1 {\sqrt{2}V_s|t|}\right)^{2/3},
\label{gksoundsound}
\end{align}
with $\Gamma_E$ Euler's gamma function\footnote{Similar expressions may be obtained for the wave number dependent diffusion coefficient and for the current-current time correlation function of the fluctuating Burgers equation.}and $V_{s'}=W_{\sigma}^{-\sigma-\sigma}$.

The leading higher order corrections are obtained by replacing in the diagrammatic expansion of the memory kernel just one pair of vertices of type $V_{\sigma}^{\sigma\sigma}$ by vertices of type $V_{\sigma}^{-\sigma-\sigma}$ or $V_{\sigma}^{HH}$. One easily shows that all these terms add contributions proportional to $|k|^{-1/3}$ to $\Gamma(k)$ and contributions proportional to $t^{-7/9}$ to the current-current correlation function. Since there are infinitely many such contributions, there seems to be no straightforward way of determining the coefficients exactly. However, estimates based on the simplest contributing diagrams can be made\cite{tbp}. Further corrections obtain from terms with $4,6,\cdots$ vertices of type $V_{\sigma}^{-\sigma-\sigma}$ or $V_{\sigma}^{HH}$. Each of these appears to be of the form $Ck^{-\mu}$ for $\Gamma(k)$ and $Dt^{-\nu}$ for the current correlation function, with $C$ and $D$ constants and $\mu$ and $\nu$ of the form $\mu=1/3-\sum_{j=2}^{\infty}m_j(2/3)^j$ and $\nu=2/3+\sum_{j=2}^{\infty}2n_j(2/3)^j$ respectively, with $m_j$ and $n_j$ natural numbers. Again, for each exponent there is an infinity of contributing terms.

The leading long time behavior of $\hat{S}_{H}(k,t)$ is determined in similar way by the sum of all contributions to $\hat{M}^{H}(k,t)$ where the first and last vertex are of type $V_H^{\sigma\sigma}$ and all other vertices are  of type $V_{\sigma}^{\sigma\sigma}$, all with the same value of $\sigma$. These terms do contain an oscillating factor $\exp(-i\sigma c_0kt)$, but these oscillations are much slower than the oscillations in any of the other terms. Since we have to include the contributions to $\hat{M}^{H}$ of either sign of $\sigma$, we cannot express $\hat{S}_{H}$ directly in terms of the Pr\"ahofer-Spohn scaling functions, but we can do so immediately for the memory kernel. A simple analysis yields to leading order
\begin{align}
\hat{M}_{H}(k,t)=2{V_{H}^2}\cos(\sigma c_0kt)
\hat{M}_{PS}(k,\sqrt{2}V_s t) ,
\label{Mheat}
\end{align}
with $V_{H}=|W_{H}^{\sigma\sigma}|$. For the $k$-dependent heat conduction coefficient and the heat current time correlation function this leads to the expressions
\begin{align}
\lambda(k)&=nC_p D_T(k)=nC_p\frac{2.1056}{2} \frac{V_H^2}{V_s}\left( \frac{V_s}{2c_0|k|}\right)^{1/3}, \\
\frac 1 L &\langle \hat{J}_{H}(0,t)\hat{J}_{H}(0,0)\rangle=
\frac{nC_p}{k_B\beta^2}{V_H}^2 \frac{2.1056}{\sqrt{3}\Gamma_E(1/3)}\left(\frac1 {\sqrt{2}V_s|t|}\right)^{2/3}.
\label{heatheat}
\end{align}

The heat mode correlation function in Fourier representation is given to leading order by
\begin{align}
&\tilde{S}_{H}(k,\omega)=
 \frac 1{-i\omega+2k^2 V_H^2\sum_{\sigma}\tilde{M}_{PS}\left(k,\frac{\omega-\sigma c_0k}{\sqrt{2}V_s}\right) } + cc.
\end{align}
From this expression and the asymptotic forms of the Pr\"ahofer-Spohn scaling functions\cite{PS} one immediately finds the long-time behavior of the heat-heat correlation function for fixed $k$ as
\begin{align}
\hat{S}_{H}(k,t)=\exp\left[-
k^2D_T(k)|t| \right].
\label{hm}
\end{align}
Higher order corrections may be obtained in similar way as for the sound modes.

The dynamic structure factor $\tilde{S}(k,\omega)$, i.e.\ the spatio-temporal Fourier transform of the density-density time correlation function exhibits Brillouin peaks at $\omega=\pm c_0k$ and a central Rayleigh peak, as usual, but, due to the anomalous transport the width and inverse height 
of the Brillouin peaks scale with $k$ as $|k|^{3/2}$\cite{delfini,delfini2} and those of the Rayleigh peak with $|k|^{5/3}$\footnote{This corresponds to the well-known value $\alpha=1/3$ for the size dependence of the heat conduction coefficient\cite{delfini,delfini2,leedadswell,rama,grassberger}}, in contrast to the usual scaling with $k^2$. Also, the shape of these peaks is not Lorentzian, but is given through the Pr\"ahofer-Spohn scaling functions as
\begin{align}
&\tilde{S}(k,\omega)=\sum_{\sigma}(\hat{n}(0),\hat{a}_{\sigma}(0))^2\tilde{S}_{\sigma}(k,\omega)\nonumber\\
&\ \ \ +(\hat{n}(0),\hat{a}_{H}(0))^2\tilde{S}_{H}(k,\omega), 
\end{align}
with 
$\tilde{S}_{\sigma}(k,\omega)$ the Fourier transform of ${S}_{\sigma}(k,t)$.

All results contained in Eqs.\ (\ref{hatfPS}-\ref{hm}) hold in the limit $L\to\infty$. For finite periodic systems they apply for $k=2\pi n/L$ and for times or inverse frequencies small compared to the sound mode traversal time $L/c_0$. For correlation functions at $k=0$, like in Eqs.\ (\ref{gksoundsound}) and (\ref{heatheat}) this time range may be extended to a value proportional to $L^{3/2}$.

It seems fair to pose that the leading long time dynamics of 1d hydrodynamic systems belong to the KPZ universality class. The sound-sound correlation functions to leading order are identical to the density-density correlation function
 of the fluctuating Burgers equation, while the heat-heat correlation functions are directly expressible in terms of KPZ memory functions in coordinate systems moving at the speed of sound. But notice that the correction terms decay only slightly faster with time and in most cases will not be negligible up to very large times.

As discussed by Pr\"ahofer and Spohn\cite{PS} a self consistent one-loop mode coupling approximation comes remarkably close to the exact solution to the Burgers equation, although there are deviations in the scaling functions of up to 10\% and not all details of the functional behavior are captured correctly. Similar results are to be expected for the one-loop mode coupling approximation to 1d hydrodynamics. Some care is required, however with using published results. In previous analyses Delfini et al.\cite{delfini} and Wang and Li\cite{lietal} assumed the sound modes were linear combinations of momentum density and displacement field (equivalent to number density in the absence of transversal modes), without contributions from the energy density. As can be seen from Eq.\ (\ref{asound}) this is justified if $(\partial p/\partial e)_n=0$ or equivalently, if $C_p=C_v$. This is the case for harmonic chains, so it will be a good approximation for weakly anharmonic chains. For general potentials corrections are needed.

It has been remarked in several places that the characteristics of heat conduction and sound dissipation change markedly under certain special conditions, such as having an anharmonic nearest neighbor potential symmetric in the deviations from the average nearest neighbor distance under zero pressure\cite{delfini,delfini2} (this includes the FPU-$\beta$ model), or zero pressure in a system of constrained hard points\cite{edhs}. This can be understood as resulting from a vanishing mode coupling amplitude $V_s$. In other words, the condition for having non-KPZ behavior quite simply and generally is $\frac n {c_0}\left(\frac{\partial c_0 }{\partial n}\right)_s=-1$. It is satisfied indeed for the classes of systems mentioned above, but in general it does not require any of the criteria quoted above, nor the condition $C_p=C_v$ as conjectured in Ref.\cite{leedadswell}. 
As discussed  by Delfini et al.\cite{delfini2},  the mode coupling under these conditions is dominated by the coupling of a sound mode to three sound modes of the same type.  Sound damping becomes almost normal. In contrast to what is stated in Ref.\cite{delfini2} it is still superdiffusive, but only logarithmically so (see Ref.\cite{Devillard} for the equivalent case of a growth model with leading nonlinearity of cubic order). The $k$-dependent heat conduction coefficient diverges roughly as $|k|^{-1/2}$ and the heat current time correlation function  decays as $t^{-1/2}$, both up to logarithmic corrections. The latter must be responsible for the gradual increase with time of the exponent $\delta$ for heat conduction from roughly 2/5 to 1/2, which has been reported for simulation results\cite{delfini2,leedadswell}.

In stationary states the $k^{\alpha-2}$ behavior of the Rayleigh peak implies a nonlinear temperature profile with, for large systems a cusp of form $|x-x_0|^{1-\alpha}$ (the inverse Fourier transform) near a boundary located at $x_0$. Such nonlinear profiles have been observed regularly in simulations, but so far I have nowhere seen mention of this simple interpretation.

Like in the case of the fluctuating Burgers equation, the mode coupling equations can also be obtained from fluctuating nonlinear hydrodynamic equations\cite{bm}. Since their structure remains exactly the same, all the results obtained above hold for all one-dimensional systems satisfying the usual Landau-Lifshitz fluctuating hydrodynamic equations with {\em finite} transport coefficients. The asymptotic long time behavior of density-density and current-current time correlation functions is independent of the values of the transport coefficients, again like for the fluctuating Burgers equation. Obviously, transport coefficients in nonlinear transport equations may be finite even if the corresponding Green-Kubo integrals are divergent. Whether this actually will happen for one and two dimensional Hamiltonian systems, as far as I know is an open question.

More detailed derivations of the results presented here will be published elsewhere. Besides comparisons to existing numerical results, new Molecular Dynamics simulations will be performed. 
Applications to quantum systems will also be studied. They look feasible but will require careful consideration of all quantum aspects.

The author acknowledges the generous support of the Humboldt foundation as well as additional support by NSF Grant No. DMR 08-02120 and 
AFOSRGrant No.AF-FA49620-01-0154 of J.\ L.\ Lebowitz. He much appreciated the hospitality of the Technische Universit\"at M\"unchen, where an important part of this work was done. He especially enjoyed many clarifying discussions with Herbert Spohn.

\end{document}